\documentclass[prl,twocolumn,showpacs,superscriptaddress,preprintnumbers,amssymb]{revtex4}
\usepackage{graphicx}% Include figure files
\usepackage{dcolumn}% Align table columns on decimal point
\usepackage{bm}% bold math
\usepackage{wasysym}
\newcommand{\beq}{\begin{equation}}
\newcommand{\eeq}{\end{equation}}
\newcommand{\beqn}{\begin{eqnarray}}
\newcommand{\eeqn}{\end{eqnarray}}

\begin{document}
\title{Conventional description of Unconventional Coulomb-Crystal phase
transition \\  in three-dimensional classical O($N$) spin-ice}
\author{Cenke Xu}
\affiliation{Department of Physics, Harvard University, Cambridge,
MA 02138}
\date{\today}

\begin{abstract}

We study the phase transition between the high temperature Coulomb
phase and the low temperature staggered crystal phase in three
dimensional classical O($N$) spin-ice model. Compared with the
previously proposed CP(1) formalism on the Coulomb-crystal
transition of the classical dimer model, our description based on
constrained order parameter is more conventional, due to a
fundamental difference between the O($N$) and the dimer model. A
systematic $\epsilon = 4 - d$ and $1/N$ expansion are used to
study the universality class of the phase transition, and a stable
fixed point is found based on our calculations for large enough
$N$.

\end{abstract}
 \maketitle

\begin{figure}
\includegraphics[width=3.0in]{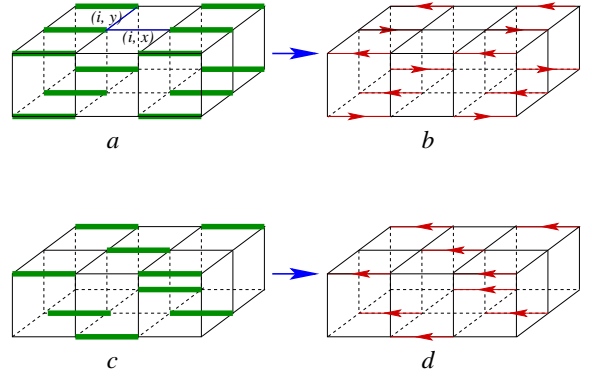}
\caption{($a$), the low temperature columnar crystalline pattern
of Eq. \ref{partition} studied in previous references; ($b$), the
magnetic field distribution corresponding to this crystalline
pattern ($a$), with a zero net magnetic field in the large scale;
($c$), the staggered dimer order; ($d$), the net nonzero magnetic
field corresponding to the dimer crystal pattern in ($c$).}
\label{dimer}
\end{figure}

The three dimensional classical dimer model (CDM), as the simplest
model with an algebraic liquid phase (usually called the Coulomb
phase), has attracted many analytical and numerical studies
\cite{sondhidimer,balentsdimer,dimersimulation1,dimersimulation2}.
The ensemble of 3d CDM is all the configurations of dimer
coverings, which are subject to a local constraint on every site:
every site is connected to precisely $n$ dimers with $0 < n < 6$
(denoted as CDM-$n$), and most studies are focused on the case
with $n = 1$. The partition function of the CDM is a summation of
all the allowed dimer coverings, with a Boltzmann weight that
favors certain types of dimer configurations. For instance, the
most standard CDM-1 takes the following form: \beqn Z &=&
\sum_{\mathrm{dimer} \ \mathrm{configuration}} \exp(-\frac{E}{T}),
\cr\cr E &=& \sum_{\mathrm{plaquettes}} - U(n_{xy} + n_{yz} +
n_{zx}). \label{partition}\eeqn $n_{xy}$, $n_{yz}$ and $n_{zx}$
are numbers defined on each unit plaquette in $xy$, $yz$ and $zx$
planes, they take value 1 when this plaquette contains two
parallel dimers, and take value 0 otherwise. When $U > 0$, the
ground state of this model favors to have as many parallel dimers
as possible, therefore when $T < T_c$, the system develops
columnar crystalline dimer patterns in Fig. \ref{dimer}$a$, while
when $T
> T_c$ the system is in the Coulomb phase with power law
correlation between dimer densities, which according to the
standard Ginzburg-Landau theory can only occur at critical points
instead of stable phases. This phase diagram has been confirmed by
a number of numerical simulations
\cite{sondhidimer,balentsdimer,dimersimulation1,dimersimulation2}.

The nature of the transition at $T_c$ attracts most efforts. If $U
> 0$, numerical studies confirm that this transition is
continuous, and the data suggest that at this transition the
discrete cubic symmetry is enlarged to an O(3) rotation symmetry
\cite{dimersimulation2}. Analytically, to describe this locally
constrained dimer system, we can introduce the ``magnetic field"
$B_\mu = n_{i, + \mu} \eta_i$, where $\eta_i = (-1)^i$ is a
staggered sign distribution on the cubic lattice. The number
$n_{i, \mu}$ is defined on each link $(i, \mu)$ (Fig.
\ref{dimer}$a$), and $n_{i,\mu} = 1, \ 0$ represents the presence
and absence of dimer. Notice that link $(i, -\mu)$ and $(i - \mu,
+\mu)$ are identical. Now the local constraint of the dimer system
can be rewritten as a Gauss law constraint $\vec{\nabla}\cdot
\vec{B}_{i} = \eta_i$. The mapping between the dimer pattern and
the magnetic field configuration is depicted in Fig. \ref{dimer}.
The standard way to solve this Gauss law constraint is to
introduce vector potential $\vec{A}$ defined on the unit
plaquettes of the cubic lattice, and $\vec{B} = \vec{\nabla}\times
\vec{A}$ \cite{sondhi2003,Hermele2005}. The background staggered
magnetic charge $\eta_i$ can be ignored in the Coulomb phase after
coarse-graining, though it plays crucial role in the crystal
phase. Since vector $\vec{A}$ is no longer subject to any
constraint, it is usually assumed that at low energy the system
can be described by a local field theory of $\vec{A}$, for
instance the field theory of the Coulomb phase reads $F \sim \int
d^3x (\vec{\nabla}\times \vec{A})^2 + \cdots$, which is invariant
under gauge transformation $\vec{A} \rightarrow \vec{A} +
\vec{\nabla}f$.

Now the Coulomb-crystal phase transition can be described by the
condensation of the matter fields which couple minimally to the
vector potential field $\vec{A}$: $\cos(\vec{\nabla}\theta^m -
\vec{A})$, and mathematically this matter field is introduced
because of the discrete nature of $\vec{A}$. Due to the staggered
background magnetic charge $\eta_i$, the matter fields move on a
nonzero background magnetic field, the band structure of the
matter fields have multiple minima in the Brillouin zone, and the
transformation between these minima encodes the information of the
lattice symmetry. For instance the transition of the CDM-1 model
is described by the CP(1) model with an enlarged SU(2) global
symmetry \cite{powell1,balentsdimer}. This field theory is highly
unconventional, in the sense that it is not formulated in terms of
physical order parameters. It is expected that more general
CDM-$n$ models can also be described by similar Higgs transition,
although the detailed lattice symmetry transformation for matter
fields would depend on $n$.

In this paper we will study an O($N$) generalization of the CDM
model. We define an O($N$) spin vector $S^a$ with unit length
$\sum_a (S^a)^2 = 1$ on each link $(i,\mu)$ of the cubic lattice,
with $a = 1 \cdots N$, and we assume that the largest term of the
Hamiltonian imposes an ice-rule constraint \cite{pauling} for
O($N$) spins on six links shared by every site: \beqn \sum_{\mu =
\pm x, \pm y, \pm z} S^a_{i,\mu} = 0. \label{onconstraint}\eeqn
Systems with this constraint is usually called the spin-ice. If $N
= 1$, each spin can only take value $\pm 1$, therefore every site
of the cubic lattice connects to three spins with $+1$ and three
spins with $-1$, which is equivalent to the CDM-3 model. Just like
the CDM, at high temperature, there is a Coulomb phase with
algebraic correlation between the O($N$) vector $S^a$, while the
low temperature crystal phase is controlled by other interactions.

In addition to the large constraint Eq. \ref{onconstraint}, we can
design the Hamiltonian as following: \beqn E &=& \sum_{i,\mu, a}
J_1 S^a_{i, -\mu}S^a_{i,\mu} + \sum_{i,a}\sum_{\mu\neq\nu} J_2
S^a_{i,\mu}S^a_{i +
\nu,\mu} %\cr\cr &+& \sum_{i}\sum_{\mu\neq\nu} J_3(\sum_a
%S^a_{i,\mu}S^a_{i, \nu})^2
. \label{model}\eeqn $J_1$ is a Heisenberg coupling between spins
along the same lattice axis, $J_2$ is a Heisenberg coupling
between spins on two parallel links across a unit square. If $J_1
> 0$ and $J_2 < 0$, in the ground state spins are antiparallel
along the same axis, but parallel between parallel links across a
unit square, which is an O($N$) analogue of the columnar state of
the CDM in Fig. \ref{dimer}$a$. If $J_1$ and $J_2$ are both
positive, in the ground state the spins are antiparallel between
nearest neighbor links on the same axis, as well as between
parallel links across a unit square, which is an analogue of the
staggered dimer configuration in Fig. \ref{dimer}$c$. Since in the
$J_1-J_2$ model Eq. \ref{model} there is no coupling between
different axes along different directions, for both cases the zero
temperature ground state of model Eq. \ref{model} has large
degeneracy, because the spins on $x$, $y$ and $z$ axes are ordered
independently, and the energy does not depend on the relative
angle between $S^a_{i,x}$, $S^a_{i,y}$ and $S^a_{i,z}$ $i.e.$ the
ground state manifold has an enlarged $[\mathrm{O}(N)]^3$
symmetry. Compared with the dimer system, the O($N$) spin vector
is not discrete, therefore we cannot use the Higgs mechanism as we
did for the dimer model, which encodes the information of the
discreteness of the dimer variables. Instead, we are going to
describe the transition with a Ginzburg-Landau (GL) theory with
order parameters. The applicability of our GL theory to the
original CDM-$n$ will be discussed later.

In this paper we will focus on the staggered spin order. Following
the magnetic field formalism of the CDM mentioned before, in order
to describe this system compactly, we introduce three flavors of
O($N$) vector field $\phi^a_\mu = S^a_{i, + \mu} \eta_i$ with $\mu
= x, y, z$, and now the constraint Eq. \ref{onconstraint} can be
rewritten concisely as \beqn \sum_\mu \nabla_\mu\phi^a_\mu = 0.
\label{phiconstraint}\eeqn Under the lattice symmetry
transformation, $\phi^a_\mu$ transforms as \beqn T_\mu &:& \ \mu
\rightarrow \mu + 1, \ \phi^a_\nu \rightarrow -\phi^a_\nu, \
(\mathrm{for} \ \mathrm{all} \ \mu, \ \nu), \cr R_{\mu, s} &:& \
\mu \rightarrow -\mu, \ \phi^a_\mu \rightarrow -\phi^a_\mu, \
\phi^a_\nu \rightarrow \phi^a_\nu, \ (\mathrm{for} \ \nu \neq
\mu), \cr R_{\mu\nu} &:& \ \mu \leftrightarrow \nu, \ \phi^a_\mu
\leftrightarrow \phi^a_\nu, \ (\mathrm{for} \ \nu \neq \mu). \eeqn
$T_\mu$ is the translation symmetry along $\mu$ direction,
$R_{\mu,s}$ is the site centered reflection symmetry, and
$R_{\mu\nu}$ is the reflection along a diagonal direction.

The mapping between $S^a_{i,\mu}$ and $\phi^a_\mu$ is very similar
to the dimer case in Fig. \ref{dimer}, and the staggered spin
order corresponds to the uniform order of $\phi^a_\mu$, and all
flavors of spin vectors are ordered. Therefore presumably
$\phi^a_\mu$ are the low energy modes close to the transition, and
we can write down the following symmetry allowed trial field
theory for $\phi^a_\mu$ with softened unit length constraint:
\beqn \mathcal{F} = \sum_{\mu, a} \phi^a_\mu (-\nabla^2 + r -
\gamma \nabla_\mu^2)\phi^a_\mu + \sum_{a} g (\sum_\mu
\nabla_\mu\phi^a_\mu)^2 + \mathcal{F}_4 \label{field}\eeqn When we
take $g \rightarrow \infty$, the constraint Eq.
\ref{phiconstraint} is effectively imposed. In Eq. \ref{field}
when $\gamma = 0$, the quadratic part of the field theory is
invariant under O($N$)$\times$O(3) transformation, the O(3)
symmetry is a combined flavor-space rotation symmetry. $\gamma$
term will break the O(3) symmetry down to the cubic lattice
symmetry. However, to the accuracy of our calculation in this
paper, the RG flow of $\gamma$ is negligible. Therefore $\gamma$
is a constant instead of a scaling function in the RG equation, we
will tentatively take $\gamma = 0$ for simplicity. In the limit
with $g \rightarrow \infty$, when $ r
> 0$ the correlation function of $\phi^a_\mu$ reads: \beqn \langle
\phi^a_\mu(\vec{q}) \phi^b_\nu(-\vec{q}) \rangle \sim
\frac{\delta_{ab}}{r+ q^2}P_{\mu\nu}, \ \ P_{\mu\nu} =
\delta_{\mu\nu} - \frac{q_\mu q_\nu}{q^2}.
\label{correlation}\eeqn $P_{\mu\nu}$ is a projection matrix that
projects a vector to the direction perpendicular to its momentum.
After Fourier transformation, this correlation function gives us
the $1/r^3$ power-law spin correlation of the Coulomb phase. When
$r < 0$, the vector $\phi^a_\mu$ is ordered.

$\mathcal{F}_4$ in Eq. \ref{field} includes all the symmetry
allowed quartic terms of $\phi^a_\mu$: \beqn \mathcal{F}_4 &=& u
\sum_\mu [\sum_a (\phi^a_\mu)^2]^2 + v \sum_{\mu \neq \nu} [\sum_a
(\phi^a_\mu)^2][\sum_b (\phi^b_\nu)^2] \cr\cr &+& w \sum_{\mu \neq
\nu} [\sum_a \phi^a_\mu\phi^a_\nu][\sum_b \phi^b_\mu\phi^b_\nu].
\label{f4}\eeqn The $u$ and $v$ terms are invariant under an
enlarged symmetry $[\mathrm{O}(N)]^3$, while the $w$ term breaks
this symmetry down to one single O($N$) symmetry plus lattice
symmetry. As already mentioned, the ground state manifold of model
Eq. \ref{model} has the same enlarged $[\mathrm{O}(N)]^3$
symmetry. However, the $w$ term can be induced with thermal
fluctuation through order-by-disorder mechanism \cite{henley1989},
or we can simply turn on such extra bi-quadratic term
energetically in the $J_1-J_2$ model Eq. \ref{model}. Just like
the $J_1-J_2$ model on the square lattice \cite{henley1989}, the
quadratic coupling $\sum_a \phi^a_\mu\phi^a_\nu$ with $\mu \neq
\nu$ as well as more complicated quartic terms like $[\sum_a
\phi^a_x\phi^a_y][\sum_b \phi^b_y\phi^b_z]$ break the reflection
symmetry of the system, and hence are forbidden.

Now a systematic renormalization group (RG) equation can be
computed with four parameters $u$, $v$, $w$ and $r$, at critical
point $r = 0$ with the correlation function Eq. \ref{correlation}.
In our calculation we will use $\epsilon = 4 - d$ expansion, and
keep the accuracy to the first order $\epsilon$ expansion. Based
on the spirit of $\epsilon$ expansion, all the loop integrals
should be evaluated at $d = 4$, and because of the flavor-space
coupling imposed by the constraint Eq. \ref{onconstraint}, we
should generalize our system to four dimension, and also increase
the flavor number to $\mu = 1 \cdots 4$. The full coupled RG
equation reads \beqn \frac{d u }{d\ln l} &=& \epsilon u - 5(8 +
N)u^2 - \frac{17 N }{4}v^2 - \frac{17 }{4}w^2 \cr\cr &-& (2 + N)uv
- \frac{17}{2}vw - 3uw, \cr\cr \frac{d v }{d\ln l} &=& \epsilon v
- \frac{2(4 + N)}{3}u^2 - \frac{52 + 37 N }{6}v^2 -
\frac{13}{6}w^2 \cr\cr &-& \frac{34(2 + N)}{3}uv - 13 vw - 14 uw,
\cr\cr \frac{d w }{d\ln l} &=& \epsilon w - \frac{8}{3}u^2 -
\frac{2}{3}v^2 - \frac{33 + 7N}{3} w^2 \cr\cr &-& 18 vw - 20 uw,
\cr\cr \frac{dr}{d\ln l} &=& 2 r - 6(2+N)ur - 9Nvr - 9wr.
\label{staggeredRG} \eeqn Solving this equation at $r = 0$, we
find stable fixed point for large enough $N$. Expanded to the
order of $\epsilon/N^2$, the stable fixed point is located at
\beqn u_\ast &=& \frac{17\epsilon}{84N} -\frac{1139\epsilon}{441
N^2} + O(\frac{\epsilon}{N^3}), \cr\cr \ v_\ast &=& -
\frac{\epsilon}{42 N} - \frac{1964\epsilon}{1323N^2} +
O(\frac{\epsilon}{N^3}), \cr\cr w_\ast &=& \frac{3\epsilon}{7N} -
\frac{4870\epsilon}{1323 N^2} + O(\frac{\epsilon}{N^3}),
\label{fixedpoint}\eeqn This expansion is valid in the limit
$\epsilon \sim 1/N^2 \ll 1$, and the fixed point values are
determined by solving the equation Eq. \ref{staggeredRG} to the
order of $\epsilon^2/N^2 \sim 1/N^6$. Close to the stable fixed
point, the three eigenvectors of the RG flow have scaling
dimensions \beqn \Delta_1 &=& - \epsilon +
O(\frac{\epsilon}{N^2}), \cr\cr \Delta_2 &=& - \epsilon +
\frac{4448\epsilon}{567N} + O(\frac{\epsilon}{N^2}), \cr\cr
\Delta_3 &=& - \epsilon + \frac{24950\epsilon}{567N} +
O(\frac{\epsilon}{N^2}). \label{dimensions}\eeqn $\Delta_3$ is the
largest scaling dimension, and according to Eq. \ref{dimensions}
the critical $N$ is $N_c = 44$. In addition to this stable fixed
point, there are seven other instable fixed points for $N > N_c$.
For instance, in the large-$N$ limit there is a fixed point at
$u_\ast = \epsilon/(24N)$, $v_\ast = \epsilon/(12N)$ and $w_\ast =
0$, which has the enlarged $\mathrm{O}(N)\times \mathrm{O}(3)$
symmetry.

\begin{figure}
\includegraphics[width=2.0in]{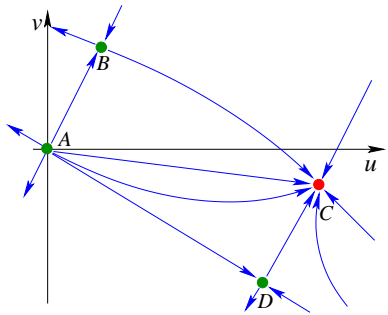}
\caption{The schematic RG flow diagram for equation Eq.
\ref{staggeredRG} in the large-$N$ limit with $w = w_\ast =
\frac{3\epsilon}{7N}$. The four fixed points are $A, (u_\ast,
v_\ast) = (0,0)$; $B, (\frac{\epsilon}{24N},
\frac{\epsilon}{12N})$; $C, (\frac{17\epsilon}{84N},
-\frac{\epsilon}{42N})$; and $D, (\frac{9\epsilon}{56N},
-\frac{3\epsilon}{28N})$. $C$ is the stable fixed point.}
\label{rg}
\end{figure}

In the large-$N$ limit, the RG equation for $w$ is decoupled from
$u$ and $v$, hence four of the eight fixed points have $w_\ast =
0$, and all the others have $w_\ast = 3/(7N)$. If we take $w =
w_\ast = 3/(7N)$, the RG flow diagram for $u$ and $v$ in the
large-$N$ limit is depicted in Fig. \ref{rg}. At the stable fixed
point Eq. \ref{fixedpoint}, $r$ is the only relevant perturbation,
with scaling dimension \beqn [r] = \frac{1}{\nu} = 2 - \epsilon +
\frac{158\epsilon}{7N} + O(\frac{\epsilon}{N^2}, \epsilon^2) \eeqn
Since at the ground state all three flavors of spin vectors are
ordered, in the field theory $\mathcal{F}_4$, $v$ should be
smaller than $2u$, which is well consistent with the stable fixed
point in Eq. \ref{fixedpoint} with negative $v_\ast$. This fixed
point has positive $w_\ast$, which favors noncollinear alignment
between spins on different axes. Therefore the transition between
Coulomb and noncollinear staggered state has a better chance to be
described by this fixed point.

Notice that had we included the anisotropic velocity $\gamma$ into
account, its leading RG flow will be at order of $\epsilon^2/N
\sim 1/N^5$, and the flow of $\gamma$ will contribute to the RG
flow of $u$, $v$ and $w$ at order of $\epsilon^3/N^2 \sim 1/N^8$,
therefore it is justified to take $\gamma$ a constant in our
calculation. When $\gamma$ is nonzero but small, the RG flows will
only change quantitatively. For instance, expanded to the first
order of $\gamma$, the scaling dimensions of the three
eigenvectors of the RG equation at the stable fixed point become
$\Delta_1 = - \epsilon$, $\Delta_2 = - \epsilon +
\frac{4448\epsilon}{567N} +
\frac{2849936\epsilon\gamma}{3988845}$, $\Delta_3 = - \epsilon +
\frac{24950\epsilon}{567N} - \frac{30088528 \epsilon
\gamma}{27921915N}$, and the scaling dimension of $r$ becomes $[r]
= \frac{1}{\nu} = 2 - \epsilon + \frac{158\epsilon}{7N} -
\frac{1032 \epsilon\gamma }{1715 N}$.

If we take $N = 1$, the $v$ and $w$ terms are identical. In this
case in addition to the trivial Gaussian fixed point, there is
only one other fixed point at $v_\ast = 2u_\ast = \epsilon/34$
with O(3) flavor-space combined rotation symmetry, which is the
same fixed point as the ferromagnetic transition with dipolar
interaction \cite{dipolar1,dipolar2}. In 3d space, the dipolar
interaction also projects a spin wave to its transverse direction.
The dipolar fixed point is instable against the O(3) to cubic
symmetry breaking, therefore when $N = 1$ our first order
$\epsilon$ expansion predicts a first order transition. We already
mentioned that the case with $N = 1$ is equivalent to the CDM-3.
However, in our GL formalism, in the ordered phase, the power law
spin-spin correlation still persists if the long range correlation
is subtracted. For instance, the fluctuation $\delta\phi_\mu =
\phi_\mu - \langle \phi_\mu \rangle$ is still subject to the
constraint $ \sum_\mu \nabla_\mu \delta \phi_\mu = 0$, therefore
although the fluctuation is gapped, it still leads to the $1/r^3$
power-law correlation. But in CDM, the ordered phase only has
short range connected dimer correlation on top of the long range
order \cite{balentsnote}. This difference is due to the fact that
our formalism does not encode the discreteness of the dimers.
Therefore one possible scenario for the CDM-3 with staggered
ground state is that, if we lower the temperature from the Coulomb
phase, after the first order transition of $\phi^a$, there has to
be another ``Higgs" like phase transition that destroys the
power-law connected correlation. Or there can be one single strong
first order transition that connects the Coulomb phase and
staggered dimer crystal directly.

The $J_1 - J_2$ model Eq. \ref{model} is invariant under cubic
symmetry transformation. One can turn on various types of cubic
symmetry breaking anisotropy to this model, and the energy favors
one of the three flavors of vector field $\phi^a_\mu$ to order.
Let us assume $\phi^a_z$ is ordered at low temperature. To
describe this transition we can turn on an extra mass gap $m$ for
$\phi^a_x$ and $\phi^a_y$ in Eq. \ref{field}, and by taking the
limit $g \rightarrow \infty$, at the critical point $r = 0$ the
renormalized correlation function for $\phi^a_z$ becomes \beqn
\langle \phi^a_z(\vec{q})\phi^b_z(-\vec{q}) \rangle \sim
\frac{\delta_{ab}}{m\frac{q_z^2}{q_x^2+q_y^2} + q_x^2+q_y^2 +
\cdots}. \eeqn In this correlation function the scaling dimension
$[q_z] = 2[q_x] = 2[q_y] = 2$, therefore this transition is
effectively a $z = 2$ transition, and $\mathcal{F}_4$ is marginal
according to power-counting. This cubic symmetry breaking
situation has been studied in Ref. \cite{anisotropy}.

In this work we studied the transition between high temperature
Coulomb phase and the low temperature staggered spin ordered phase
in the O($N$) spin-ice model. Higher order $\epsilon$ and $1/N$
expansion are demanded to obtain more quantitatively accurate
results. The model Eq. \ref{model} can be simulated directly
numerically, and our RG calculation can be tested. The columnar
phase with $J_1
> 0$ and $J_2 <0$ is also interesting. But since the columnar order
does not correspond to the uniform order of $\phi^a_\mu$, the
order parameter description is more complicated. We will study
this situation in future.

The CDM is also considered as a simple analogue of the spin-ice
materials such as $\mathrm{Ho_2Ti_2O_7}$ and
$\mathrm{Dy_2Ti_2O_7}$
\cite{andersonspinice,sondhispinice1,sondhispinice2}, where the
$\mathrm{Ho^{3+}}$ and $\mathrm{Dy^{3+}}$ magnetic moments reside
on the sites of a pyrochlore lattice, and the ground state of
these moments satisfies the same ice-rule constraint as Eq.
\ref{onconstraint}. The Coulomb phase of the spin ice materials
with fractionalized ``monopole" like defect excitation has been
observed experimentally \cite{spinicemonopole1}. The formalism
developed in our work is largely applicable to the pyrochlore
lattice, while the symmetry analysis and the number of quartic
terms are different. A complete symmetry analysis is demanded in
order to correctly understand the O($N$) spin-ice on the
pyrochlore lattice.

The author thanks very helpful discussion with Leon Balents, Subir
Sachdev and T. Senthil. This work is sponsored by the Society of
Fellows, Harvard University.

\bibliography{dimer}

\end{document}